# Multi-objective optimization of actuation waveform for high-precision drop-on-demand inkjet printing

Hanzhi Wang[1], Yosuke Hasegawa[2,*]

1. Department of Mechanical Engieering, The University of Tokyo, 4-6-1 Komaba, Meguro-ku, Tokyo 153-8505, Japan

2. Institute of Industrial Science, The University of Tokyo, 4-6-1 Komaba, Meguro-ku, Tokyo 153-8505, Japan

**ABSTRACT:** Drop-on-demand (DOD) inkjet printing has been considered as one of promising technologies for the fabrication of advanced functional materials. For a DOD printer, high-precision dispensing techniques for achieving satellite-free smaller droplets, have long been desired for patterning thin-film structures. Optimization of actuation waveform driving a DOD inkjet printer is one of the most versatile and effective strategies to obtain high-precision droplets. Considering the complexity of physics behind the droplet dispensing mechanisms and the large degrees of freedom in the applied waveforms, conventional trial-and-error approaches are not effective for searching the optimal waveform. The present study considers the inlet velocity of a liquid chamber located upstream of a dispensing nozzle as a control variable and aims to optimize its waveform using a sample-efficient Bayesian optimization algorithm. Firstly, the droplet dispensing dynamics are numerically reproduced by using an open-source OpenFOAM solver, *interFoam*, and the results are passed on to another code based on pyFoam. Then, the parameters characterizing the actuation waveform driving a DOD printer are determined by the Bayesian optimization (BO) algorithm so as to maximize a prescribed multi-objective function expressed as the sum of two factors, i.e., the size of a primary droplet and the presence of satellite droplets. The results show that the present BO algorithm can successfully find high-precision dispensing waveforms within 150 simulations. Specifically, satellite droplets can be effectively eliminated and the droplet diameter can be significantly reduced to 24.9% of the nozzle diameter by applying the optimal waveform. Through a detailed comparison of three types of waveforms, "pull-push" waveforms are effective for smaller droplet dispensing. Moreover, the prediction using the Gaussian process regression (GPR) suggests that the size of the primal droplet is highly correlated with the period of a waveform.

Finally, the criterion for achieving single-droplet dispensing is proposed based on the energy budget analysis.

## I. INTRODUCTION

Drop-on-demand (DOD) inkjet printing is a high-precision and cost-effective technology for material delivering and structure patterning[1] and thus attracts much attention as one of the promising technologies for the fabrication of functional materials, such as thin-film transistors,[2,3,4] electrochemical devices,[5] electronics,[6,7] solar cells,[8,9,10] sensors,[11] bio-tissues,[12,13,14] to name a few. Typically, high-resolution material structures are crucial for improving the performance of inkjet-printed functional materials, especially for thin-film structures. To achieve this goal, a high-performance inkjet printer has long been desired for accurately locating materials with high resolution.

DOD inkjet printers deliver materials onto the substrate by ejecting small droplets, and its printing resolution mainly depends on the droplet size and the positioning accuracy. Generally, the dispensed droplet size is limited by the nozzle orifice diameter. However, simply decreasing a nozzle size does not always guarantee high-resolution dispensing, since a smaller nozzle tends to suffer from clogging.[15] Therefore, achieving a smaller droplet without reducing the nozzle diameter is desirable. Moreover, another important factor that can significantly affects dispensing resolution is the emergence of satellite droplets accompanying a primary droplet, since they are often printed on undesired locations. Hence, reducing the size of the primary droplet and preventing the emergence of satellite droplets are two major technical challenges for achieving high-precision inkjet printing.

The inkjet printing processes consist of three subsequent physical processes, namely, ink-thread ejection, pinch-off and thread contraction.[16] After the pinch-off stage, the ink-thread ejected from a nozzle could contract into a single-droplet or breakup into a primary droplet and single/multiple satellites depending on a printing condition. The printability[17,18] of a single-droplet is often characterized by a non-dimensional number $Z$ ($Z=\sqrt{\rho_l \sigma d_n/(2\mu_l^2)}$),[19] which is defined based on the nozzle diameter $d_n$, and ink properties such as the density $\rho_l$, the surface tension $\sigma$, and the viscosity $\mu_l$ of the ink. Derby and Reis[20] systematically analyzed the influence of ink properties on the dispensing results and concluded that $Z$ number must lie within the range of $1<Z<10$ for stable droplet formation. Since then, several researchers have also attempted to define the printable regions using $Z$ number through experimental studies and suggested similar, but different conditions for single-droplet dispensing, i.e., 2.8< $Z$ <10,[21] 1.4< $Z$ <14,[22] and 0.7< $Z$ < 42.[23] In these existing studies, the suggested printable regions are somehow consistent for the lower limit of $Z$ number, while the upper limit shows wide variation. Generally, at a low value of $Z$, viscous effects are dominant and prevent droplet ejection from the nozzle, whereas at a high value of $Z$ a primary droplet often splits into multiple satellite droplets. The disagreement on the upper limit of printable $Z$ indicates that the generation of satellite droplets is determined not only by the geometric nozzle diameter and the physical properties of the ink, but also by fluid dynamical factors. Actually, the dimensionless number $Z$ does not include ink velocity, so that the inertia effects are not explicitly taken into account. Accordingly, with appropriate control of dispensing processes, there still exist the possibilities for eliminating the satellite droplets even at higher $Z$ numbers.[24]

One of the most effective ways of controlling dispensing processes is to optimize an actuation waveform.[15,25,26] This approach is not limited to a particular nozzle, ink properties, and printing principles, and therefore is promising for wide aplications. In the past decades, optimization of the actuation waveform have been attempted. For example, Aqeel et al.[27] conducted a parametric study on the effects of the actuation waveform for reducing the droplet size. Unipolar and bipolar waveforms were investigated for three different ink properties. The droplet diameter was successfully reduced to 40.0% of the nozzle diameter. Gan et al.[28] investigated different types of waveforms for droplet size reduction. By tuning waveform parameter, they could achieve a single-droplet as small as 41.5% of the nozzle diameter. Ning et al.[29] considered three types of the actuation waveforms for achieving satellite-free dispensing. The morphological properties of printed silver electrodes were greatly improved with a proper choice of a waveform for eliminating satellites. Shin et al.[30] applied double waveforms to control the droplet formation for a low viscosity fluid. By changing the time separation between the pulses, a single droplet could be easily dispensed. Oktavianty et al.[31] improved the quality of inkjet printing through tuning an actuation waveform. Satellite-free droplets with a wide range of droplet sizes were achieved by optimizing "U-type" and "W-type" waveforms. Further, Snyder et al.[24] proposed an automatic waveform tuning strategy by combining genetic algorithm (GA) and real-time image processing. The drop diameter could be reduced to as low as 25.6% of the nozzle diameter with sufficiently high droplet velocity to ensure the positional accuracy of droplet dispensing.

From these existing studies, there is no doubt that the droplet dispensing processes can be greatly improved by modifying an actuation waveform so that a single or smaller droplet can be obtained. However, most existing studies consider only one of these two objectives, and therefore the waveform characteristics for simultaneously achieving the two objectives have not been clarified yet. Moreover, the existing optimization studies commonly predetermined the functional form of a driving signal for an actuator, and only a few parameters characterizing the prescribed function have been optimized mostly through a parametric survey to achieve a single objective. Considering the complexity of the droplet dispensing mechanisms and the large-degrees of freedom of a waveform, a conventional trial-and-error approach is not effective for exploring the optimal waveform in a large design space. Recently, learning-based methods have attracted much attention for predicting and optimizing complex printing processes. Wu and Xu[32] and Brishty et al.[33] developed learning-based predictive models to predict droplet formation processes in inkjet printing. Several parameters including ink properties and actuation waveform parameters were included to predict droplet volume, velocity and printability. The results indicate that learning-based models generally show good prediction accuracy for the complex dispensing processes. However, for learning-based predictive models, the selection of learning algorithms is crucial for the prediction accuracy.[32] Moreover, the prediction accuracy also highly depends on the size and completeness of training datasets. Bayesian optimization (BO) is a robust global optimization algorithm combining prediction models, typically using Gaussian process regression (GPR) model, and acquisition functions. This method does not require large training datasets and can be used to efficiently optimize complex printing processes within a few experiments, which could significantly save experimental time and cost.[34]

The present study focuses on the waveform optimization of a DOD inkjet printer using a sampling-efficient BO algorithm. Two multi-objective functions based on the ink-thread properties at the pinch-off and near-spherical stages, respectively, are proposed. Then, the BO algorithm is applied to optimize parameters characterizing an actuation waveform for eliminating satellite droplets and reducing the size of a primary droplet simultaneously. By leveraging the unique property of a Gaussian process regression (GPR) model employed in BO, the waveform characteristics for achieving single and smaller droplet dispensing are predicted. Furthermore, the criterion of the applied waveform for obtaining a single droplet is developed based on the energy budget analysis of the training samples.

The rest of this paper is organized as follows. In Sec. II, the numerical algorithms for the simulation of droplet dispensing and also optimization of an actuation waveform are introduced. In Sec. III, the current numerical method is validated through the comparison with experimental data from an existing literature.[35] In Sec. IV, the optimization results are presented and the waveform characteristics for realizing small single-droplet dispensing are discussed. The optimal wave period is also predicted using the GPR models trained by the present numerical results. Moreover, the physical mechanisms of achieving a single-droplet dispensing are discussed through energy budget analysis. Finally, the present study is summarized in Sec. V.

## II. METHODS

The present optimization study is based on numerical simulation and the schematic of the overall optimization procedures are illustrated in Fig. 1. As shown in Fig. 1, the current optimization framework includes 1) waveform generation using a *Bezier* curve, 2) CFD simulation based on OpenFOAM and PyFoam package, 3) evaluation of an objective function, and 4) Bayesian optimization of control variables. The entire optimization loop is controlled by a Python script. The detailed procedures of each part will be described in the following sections.

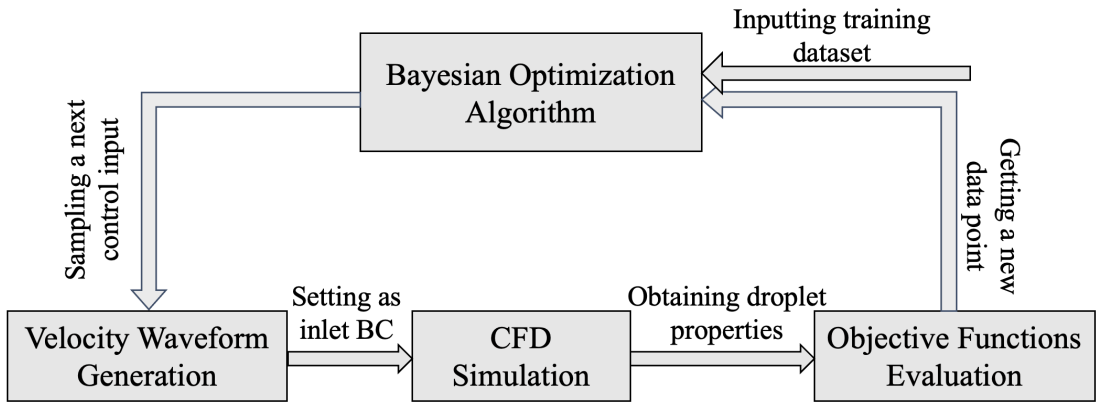

FIG. 1. Schematic diagram of the overall optimization process.

### A. CFD model

The computational domain considered in the present study is illustrated in Fig. 2. The vertical axis is denoted by $z$, while the horizontal direction is $r$. The origin is located at the center of the nozzle exit. Since the Reynolds number is generally small, flow is assumed to be laminar and axisymmetric with respect to the $z$ axis. The present geometry is determined by the experimental settings of Castrejón-Pita et al.[35] where a nozzle diameter of 2 mm is employed. We have checked that the present domain size is large enough for reproducing the whole droplet dispensing process and further extension of the computational domain does not alter the present results and conclusions.

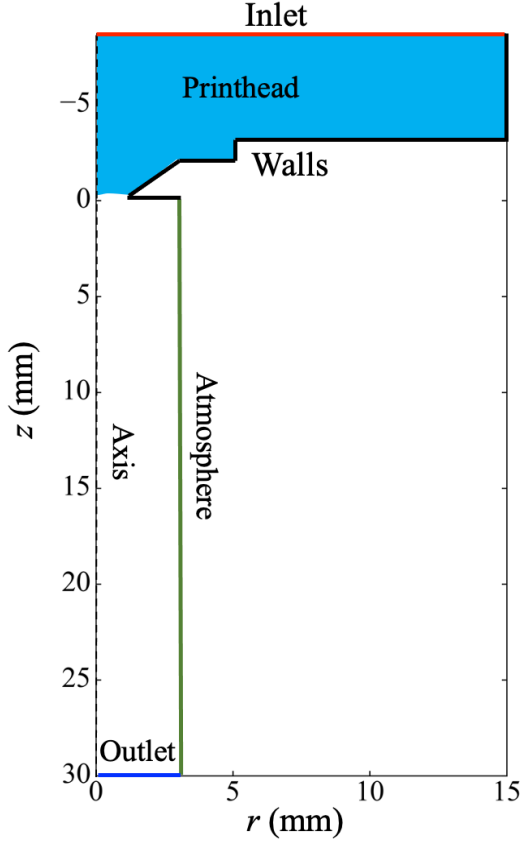

FIG. 2. Schematic of the 2D-axisymmetric computational domain.

We use an open-source solver, *interFoam*,[36] for simulating the droplet dispensing process. The *interFoam* solver uses the volume of fluid (VOF) method[37] for capturing a gas-liquid interface and solving two-phase flow in a unified manner. Both gas and liquid are assumed as Newtonian fluids and their temperatures are identical and constant. We also assume that both fluids are incompressible and laminar. Accordingly, the unified governing equations for gas and liquid can be written as follows:

$$\nabla \cdot \boldsymbol{u} = 0, \tag{1}$$

$$\rho \frac{D\boldsymbol{u}}{Dt} = -\nabla p + \mu \nabla^2 \boldsymbol{u} + \rho \boldsymbol{f}, \tag{2}$$

where $\boldsymbol{f}$ is the surface tension force, which is given by

$$\boldsymbol{f} = \sigma \kappa \nabla \alpha, \tag{3}$$

where $\sigma$ is the surface tension (N m$^{-2}$), $\alpha$ is the liquid volume fraction, and $\kappa$ is the gas-liquid surface curvature. The evolution equation for the liquid volume fraction $\alpha$ in the VOF framework is given by

$$\frac{\partial \alpha}{\partial t} + (\boldsymbol{u} \cdot \nabla)\alpha = 0. \tag{4}$$

The fluid density $\rho$ and dynamic viscosity $\mu$ are smoothly switched between the corresponding values for gas and liquid liquid based on the phase fraction $\alpha$ as follows.

$$\rho = \alpha \rho_l + (1 - \alpha)\rho_g, \quad (5)$$

$$\mu = \alpha \mu_l + (1 - \alpha)\mu_g, \quad (6)$$

where the subscripts $l$ and $g$ denote the values in the liquid and gas phases, respectively.

Taking the nozzle radius ($d_n/2$) and the droplet ejection speed ($U$) as the typical length and velocity scales[20,24] the non-dimensional numbers including Reynolds number ($Re$) and Weber number ($We$) based on the physical properties of the liquid can be defined as follows:

$$Re = \frac{\rho_l U d_n}{2\mu_l}, \quad (7)$$

$$We = \frac{\rho_l U^2 d_n}{2\sigma}. \quad (8)$$

We also note that these dimensionless number can be related to the non-dimensional number $Z$ as $Z = Re/We^{1/2} = [\rho_l \sigma d_n/(2\mu_l^2)]^{1/2}$.[20,24]

In the present study, the gas phase is always considered as air and its physical properties are used for the simulation. For the liquid phase, glycerol/water mixture[35] is used for the validation, and ethylene glycol (EG)/water mixture[22] with modified viscosities are also considered for the optimization study. The details of the ink properties and the corresponding $Z$ numbers considered in the present study are summarized in Table I.

TABLE I. Ink properties.[22,35]

| Inks | Density (kg m$^{-3}$) | Viscosity (mPa s) | Surface tension (N m$^{-2}$) | $Z$ number |
|---|---|---|---|---|
| Glycerol(0.85)/water(0.15) | 1222 | 100 | 0.064 | 2.8 |
| EG(0.25)/water(0.75) | 1014 | 27.2* | 0.067 | 9.6 |
| EG(0.25)/water(0.75) | 1014 | 5.44* | 0.067 | 48 |

*Viscosities are modified to obtain target $Z$ numbers.

The velocity, pressure and liquid volume fraction boundary conditions used for the current calculation are given in Table II. It should be noted that we give a time dependent, and spatially uniform inlet velocity $u_z(t)$ at $z = -8.6$ mm as an actuation waveform, which will be optimized using BO algorithm. We also note that the coordinate $z$ is pointing downward in Fig. 1, so that the positive $u_z(t)$ indicates pushing the liquid out of the nozzle downward, while the negative value corresponds to pulling the liquid into the chamber upward. The parameterization of the actuation waveform is achieved by using a cubic *Bezier* curve, and the detailed procedures will be explained in the Sec. II B.

A no-slip condition is applied to the wall boundary, and the wettability of the nozzle wall is taken into account by imposing a constant contact angle $\theta_{ca}$, which provides a boundary condition for the unit normal vector of the liquid phase fraction, $\boldsymbol{n}_\alpha = \nabla\alpha/|\nabla\alpha|$, at the contact line region as follows:[38]

$$\boldsymbol{n}_\alpha = \boldsymbol{n}_w \cos\theta_{ca} + \boldsymbol{n}_t \sin\theta_{ca}, \tag{9}$$

where $\boldsymbol{n}_w$ is the unit wall normal vector, and $\boldsymbol{n}_t$ is the unit wall tangential vector perpendicular to the contact line. The contact angle is only computed at the cells next to the wall at the contact line. Typically, the value of contact angle is determined from the nozzle surface and ink properties. In the present study, a contact angle of 60° is assumed according to the water contact angle of a polymethyl methacrylate nozzle used in the reference experimental study.[39]

The pressure-implicit with splitting of operators (PISO) scheme is used for the pressure-velocity coupling. For time advancement, the first order Euler implicit scheme is used, while the second-order Gauss scheme is chosen for the spatial discretization of the advection and diffusion terms. A block mesh with refined meshes near the *z* axis is used for simulation. After the grid convergence checking, a mesh with 46520 cells is used for the present simulation. The calculation time step is set to be adjustable by restricting the maximum Courant number to 0.1.

TABLE II. Boundary conditions for the present computational domain.

| **Boundaries** | *p* | $\boldsymbol{u}$ | $\alpha$ |
| --- | --- | --- | --- |
| **Inlet** | $\frac{\partial p}{\partial n} = 0$ | $u_r = 0$, $u_z = u_z(t)$ | $\alpha = 1$ |
| **Outlet** | $p + 0.5|\boldsymbol{u}|^2 = 0$ | $\frac{\partial u_i}{\partial n} = 0$, $u_r = 0$ | $\frac{\partial \alpha}{\partial n} = 0$ |
| **Walls** | $\frac{\partial p}{\partial n} = 0$ | $u_i = 0$ | $\theta_{ca} = 60°$ |
| **Atmosphere** | $p + 0.5|\boldsymbol{u}|^2 = 0$ | $\frac{\partial u_i}{\partial n} = 0$, $u_z = 0$ | $\alpha = 0$ |

**B. Actuation waveform**

The waveform of the liquid velocity at the inlet of the liquid chamber is generated using a *Bezier* curve which is a parametric curve controlled by a set of control points. In the present study, four control points are considered. The equation of a cubic *Bezier* curve is defined as follows:

$$B\{(t_i, u_{z,i})|_{i=1,2,\ldots,n}\} = \sum_{i=0}^{n} \binom{n}{i}(1-P)^{n-i}P^i(t, u_{z,i}), \qquad 0 \leq P \leq 1. \qquad (10)$$

Points $(t_0, u_{z,0})$ and $(t_3, u_{z,3})$ determine the starting and ending points, whereas points $(t_1, u_{z,1})$ and $(t_2, u_{z,2})$ control the shape and amplitude of the waveform (see, in Fig. 3). In addition, the first point is fixed to $(t_0, u_{z,0}) = (0, 0)$, and also the liquid velocity at the end point is fixed to zero, i.e., $u_{z,4} = 0$. Therefore, the rest of five parameters, i.e., $(t_1, t_2, t_3, u_{z,1}, u_{z,2})$, are control variables to be optimized in the present study.

In Fig. 3, three typical waveform profiles, namely, unipolar, bipolar I and bipolar II, are presented with corresponding four control points. The unipolar waveforms are characterized with a "push" velocity profile, whereas the bipolar waveforms have "push-pull" and "pull-push" velocity profiles for types I and II, respectively. One of the advantages using the *Bezier* curve is that these different types of waveforms can easily be expressed by arranging the control points.

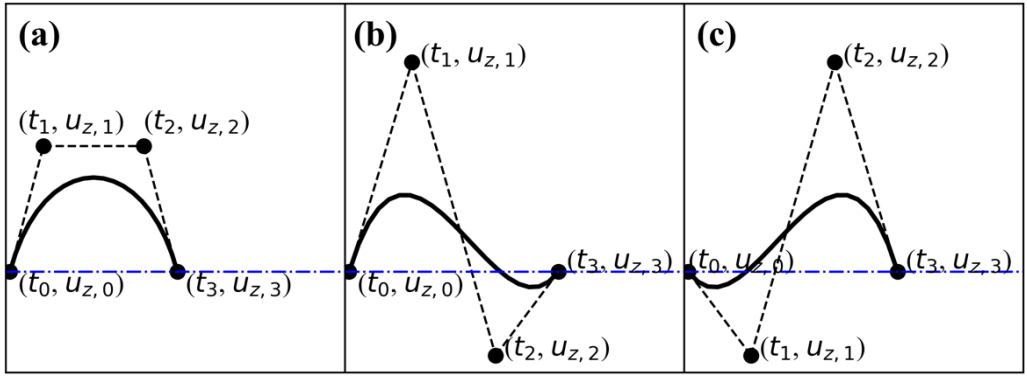

FIG. 3. Illustration of waveform generation using a cubic *Bezier* curve. (a) Unipolar waveform with $u_{z,1} > 0$ and $u_{z,2} > 0$, (b) bipolar I waveform with $u_{z,1} > 0$ and $u_{z,2} < 0$, and (c) bipolar II waveform with $u_{z,1} < 0$ and $u_{z,2} > 0$.

**C. Objective functions**

The processes for DOD droplet dispensing are characterized by ejection and stretching, pinch-off, and contraction of thread into a single droplet or breakup into a primary droplet and satellite droplets.[16] The dispensing results can be categorized as single-droplet (SD), multiple-droplet (MD) and without-droplet (WD) dispensing. The main objective of this research is to dispense high-precision droplets using DOD inkjet printer, which requires a single and smaller droplet. Therefore, two objectives, i.e. a satellite-free and smaller primary droplet, are considered for the current optimization problem. As discussed below, we introduce two different multi-objective functions based on the pinch-off and near-spherical stages, respectively.

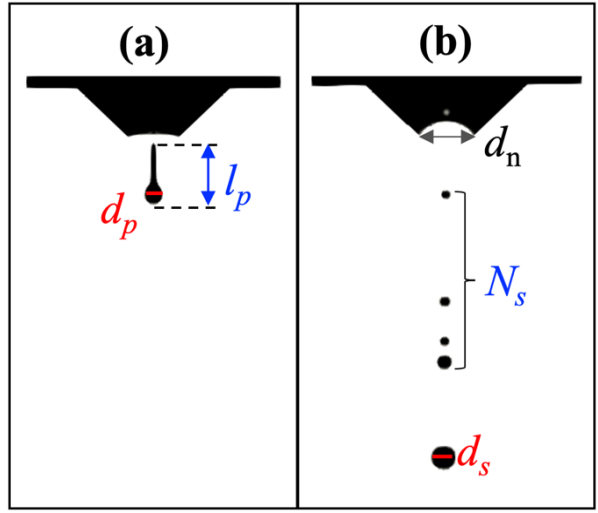

FIG. 4. Illustration of the (a) pinch-off and (b) the near-spherical stages of droplet dispensing.

**1. Multi-objective function based on the ink-thread morphology at pinch-off stage**

During droplet dispensing, the growth of the most unstable disturbance is responsible for the formation of satellite droplets. Therefore, the morphological properties of an ink-thread at the pinch-off stage are crucial for the final dispensing results. A larger ink-thread length at the pinch-off stage usually leads to a higher possibility of generating satellite droplets.[17] Therefore, Dong et al.[16] proposed a criterion for single-droplet dispensing based on the ratio of the ink-thread length $l_p$ at the pinch-off stage to the nozzle diameter $d_n$ as follows:

$$\frac{l_p}{d_n} < \frac{\zeta}{\alpha_{max}^*} + 1 = \frac{l_p^*}{d_n}, \qquad (11)$$

where the constant $\zeta$ varies from 0.9 to 1.1 based on the experimental results, $\alpha_{max}^*$ is a coefficient depending on the temporal growth rate of disturbance, while $l_p^*$ is a critical ink-thread length for the single-droplet dispensing.

Accordingly, we define the ink-thread length $l_p$ and the droplet diameter $d_p$ at the pinch-off stage as depicted in Fig. 4 (a). The first objective is to make their ratio $l_p/d_p$ smaller for reducing the possibility of satellite generation. As for the size of a primary droplet, we consider the droplet diameter normalized by the nozzle diameter, i.e., $d_p/d_n$. Consequently, we define a multi-objective function to be maximize as follows:

Multi-objective function I:

$$f_\mathrm{I}\big(B\{(t_i,u_{z,i})|_{i=0,1,2,3}\}\big) = \begin{cases} -\beta_1\left(\dfrac{l_p}{d_p}\right) + \left(\dfrac{d_n}{d_p}\right) & \ldots(SD, MD) \\ 0 & \ldots(WD) \end{cases}, \qquad (12)$$

where $\beta_1$ is a weight coefficient to adjust the relative importance of the two objectives. When the printer fails to dispense a droplet, the BO algorithm assign 0 to the objective function for penalization (see, Eq. (12)). Using higher scale factor $\beta_1$, the optimization more focus on a shorter ink-thread length which is beneficial for single-droplet dispensing, whereas setting a lower $\beta_1$ puts more emphasis on the size of a primary droplet. In the current study, the weight coefficient is set to be $\beta_1 = 0.1$ which is determined from the initial training samples to make sure that dispensing cases with a smaller primary droplet and fewer satellites have larger values of the objective function (12).

### 2. Multi-objective function based on the morphology of the ejected droplet

The multi-objective function proposed in the previous subsection is based on the morphology of the ink-thread at the pinch-off stage. Here, we introduce the second multi-objective function based on the morphology of the final droplet schematically shown in Fig. 4b. Based on these parameters, the second multi-objective function can be defined as follows.

Multi-objective function II:

$$f_\mathrm{II}\big(B\{(t_i,u_{z,i})|_{i=0,1,2,3}\}\big) = \begin{cases} -N_s^{\beta_2} + \left(\dfrac{d_n}{d_s}\right) & \ldots(SD, MD) \\ -10 & \ldots(WD) \end{cases}, \qquad (13)$$

where $d_s$ and $N_s$ are the primary droplet diameter and the number of satellite droplets at the near-spherical stage, respectively. The first term on the right side of Eq. (13) is introduced to express the penalty of satellite generation, whereas the second term quantifies the smallness of a primary droplet. The exponential form of the first term is devised to balance the two objectives. Specifically, $\beta_2 = 0.7$ is used in the present study. For failed droplet dispensing (WD) cases, the BO algorithm assigns -10 to the objective function for penalization as shown in Eq. (13).

### 3. Design parameter ranges

As discussed in Sec. II B, we consider totally 5 variables [$t_1$, $t_2$, $t_3$, $u_{z,1}$, $u_{z,2}$] as control parameters to define the velocity waveforms. The ranges of the design parameters are shown in Table III which are determined from the typical time-scale and the velocity amplitude of existing printers. Additional constrains on $t_3$, $u_{z,1}$ and $u_{z,1}$ are added during the optimization to reflect some common facts that a wave period should be positive and only 'pull' waveform obviously cannot dispense a droplet. These constrains are introduced to limit the searching domain as shown in Table III.

TABLE III. Design parameter ranges for waveform generation.

| Design parameters | Parameter ranges | Additional constrains |
|---|---|---|
| $t_1$ | 0 ~ $t_3$ (ms) | - |
| $t_2$ | 0 ~ $t_3$ (ms) | - |
| $t_3$ | 0 ~ 5 (ms) | $t_3 > 0$ |
| $u_{z,1}$ | -25 ~ 25 (mm/s) | $u_{z,1} > 0$ if $u_{z,2} < 0$ |
| $u_{z,2}$ | -25 ~ 25 (mm/s) | $u_{z,2} > 0$ if $u_{z,1} < 0$ |

**D. Bayesian optimization algorithm**

Bayesian optimization (BO) method is a sample-efficient method that can find the optimum design parameters from limited samples, and thus has been widely used for complex optimization problems.[40,41] Generally, the BO algorithm treats relevant physical processes as a black box and develops a surrogate model to relate the inputs (control variables) and the final output. In the present study, the inputs are the control points of the cubic *Bezier* curve, which dictates the inlet velocity of the liquid chamber, whereas the final output is the objective function defined by either Eq. (12) or (13).

One of the widely used surrogate models is the Gaussian distribution model. First, a probability distribution is assumed over the objective function which is known as a prior distribution. Then, given a set of observations, i.e., a training dataset, the posterior distribution can be updated based on the Bayes' theorem. This posterior distribution typically gives the best fitting to the training dataset and also provides the estimate and uncertainty of the objective function for an unseen input. The main procedures for BO are described below.

First, based on a training dataset which has *n* observations $D := \{X^*, f^*\}$, where $X^*$ represents a set of the inputs and $f^*$ is the corresponding output, we assume that the posterior probability density function (PDF) of the ouput *f* for an unseen input $X$ is a Gaussian process as follows:

$$f(X) \sim \mathcal{N}(m(X,D), cov(X,D)), \qquad (14)$$

where $m(,)$ and $cov(,)$ are the mean and covariance functions, respectively. They are obtained from the following formula:

$$m(X,D) = K(X,X^*)K(X^*,X^*)f^*, \qquad (15)$$
$$cov(X,D) = K(X,X) - K(X,X^*)K(X^*,X^*)^{-1}K(X^*,X). \qquad (16)$$

Here, $K(x_i, x_j)$ is a kernel matrix which dictates the covariance of the output $f$ at two arbitrary different inputs, $x_i$ and $x_j$. We employ one of the most commonly used kernels, *Matern 5/2* kernel, which is given below:

$$K(x_i, x_j) = \theta_0 \left(1 + \sqrt{5}\frac{d(x_i, x_j)}{l} + \frac{5}{3}\left(\frac{d(x_i, x_j)}{l}\right)^2\right)\exp\left(-\sqrt{5}\frac{d(x_i, x_j)}{l}\right), \quad (17)$$

where $d(x_i, x_j)$ is the Euclidean distance. The covariance amplitude $\theta_0$ and the length scale $l$ in Eq. (17) are two important hyperparameters describing the posterior distribution. They are dynamically learned from the sample data by maximizing the following log-marginal-likelihood function:

$$\log p(f|X) = -\frac{1}{2} f^T K^{-1} f - \frac{1}{2}\log|K| - \frac{n}{2}\log(2\pi). \quad (18)$$

After obtaining the posterior distribution, we sample a next point ($x^+$) by maximizing a prescribed acquisition function $a(x)$. In this study, the expected improvement (EI) acquisition function defined below is used:[40]

$$x^+ = argmax_{x \in X} a(x), \quad (19)$$

$$a(x) = m(x|D)[\gamma(x)\Phi(\gamma(x)) + \phi(\gamma(x))], \quad (20)$$

$$\gamma(x) = \frac{m(x|D) - f^*_{max}}{cov(x|D)}, \quad (21)$$

where $\phi(.)$ and $\Phi(.)$ are the PDF and the cumulative distribution function (CDF) of a standard normal distribution, respectively. Finally, the new set of samples are added to the training dataset and the above procedures are repeated until the optimization is completed.

## III. CFD SIMULATION AND VALIDATION

Before optimizing an actuation waveform, we first validate our simulation code by comparing the present results with existing experimental data.[35] The present printhead structure used for validation is identical to the experimental settings of Castrejón-Pita et al.[35] as shown in Fig. 2. In their experiment, the time-dependent vertical velocity inside fluid chamber is measured at z = -8.6 mm, which is the same location as inlet boundary for simulation, using laser Doppler anemometry. Experimental data shown in Fig. 5 is the superposition of five independent trials. In the present study, we construct the inlet boundary condition by fitting a *Bezier* curve to the experimental data.

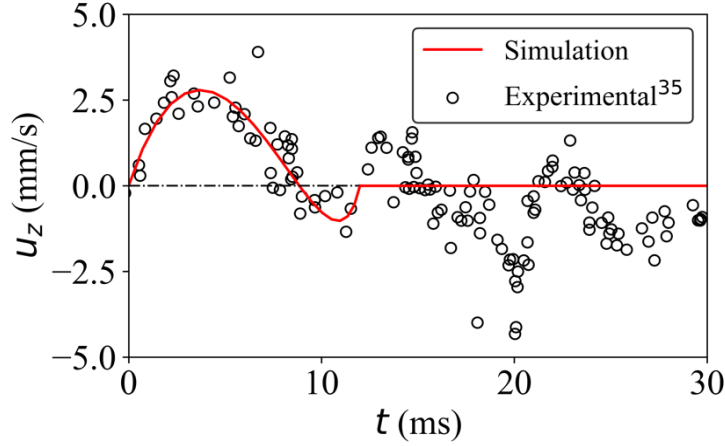

FIG. 5. Experimental data of liquid velocity in Ref. [35] and its fitting by *Bezier* curve for creating the inlet condition in the present simulation.

In Figs. 6 and 7, the time evolution of the tip position of a primary droplet and the visualization of dispensing processes obtained in the present simulation are compared with existing experimental and numerical data.[35] As shown in Fig. 6, the time evolution of the droplet tip position calculated from the present *interFoam* simulation shows good agreement with the experimental and numerical results in the previous study. Moreover, as shown in Fig. 7, the evolution of ink-thread during the dispensing process is also consistent with the previous experimental images. The good agreement between the present numerical results and those in the previous studies validates the current numerical scheme. In the next section, we will conduct optimization of the actuation waveform for minimizing the droplet size without having satellite droplets under the same printer configuration.

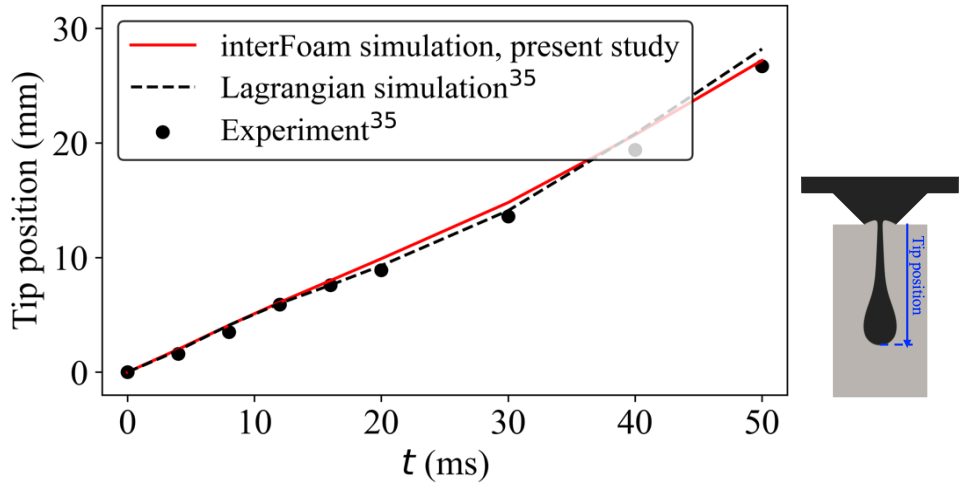

FIG. 6. Comparison of time evolution tip positions of *interFoam* simulation with the previous experimental and simulation results.[35]

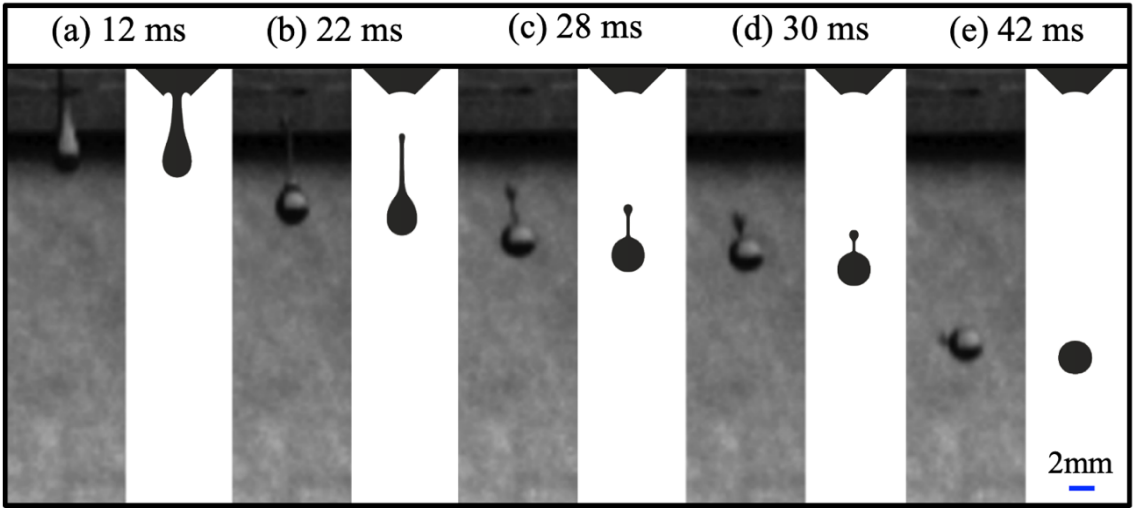

FIG. 7. Comparison of the present *interFoam* simulation results (right) with existing experimental images{https://doi.org/10.1103/PhysRevE.83.036306}[35] (left) at different time instance. Here, $t = 0$ ms for simulation is the starting point of the velocity waveform, while for experiments, it is an approximated time at which the tip position increases from zero.

**IV. OPTIMIZATION OF ACTUATION WAVEFORM**

In this section, we first present the results of multi-objective optimization using BO algorithm. Then, the detailed waveform characteristics are analyzed by comparing similar obtained waveforms but with different droplet dispensing results. Moreover, the optimal wave period is predicted leveraging the GPR model. Finally, a physical criterion based on energy budget is developed to clarify the relations between applied waveforms and dispensing results.

Before starting the optimization, 30 samples are randomly generated as an initial training dataset (TD) by using Eq. (10) with the ranges of the design parameters listed in Table III. In order to confirm the reproducibility and universality of our results, we prepare two independent randomly generated training datasets, i.e., TD1 and TD2. The resulting totally 60 samples are shown in Fig. 8.

Ink properties used in the present paper are characterized by the non-dimensional number $Z$. Typically, for a higher $Z$ number, it tends to be more difficult to avoid generation of satellites, and the upper limit of printable $Z$ suggested by Derby[19] is almost 10. Therefore, a relatively high $Z$ number of $Z = 9.6$ with a density of 1014 kg m$^{-3}$, a dynamic viscosity of 27.2 mPa s, and a surface tension of 0.067 N m$^{-2}$ is selected as ink properties for the optimization. Furthermore, in order to demonstrate the effectiveness of the present optimization approach, we also consider a case with an even larger value of $Z = 48$. The details of ink properties used in the current simulation are shown in Table I.

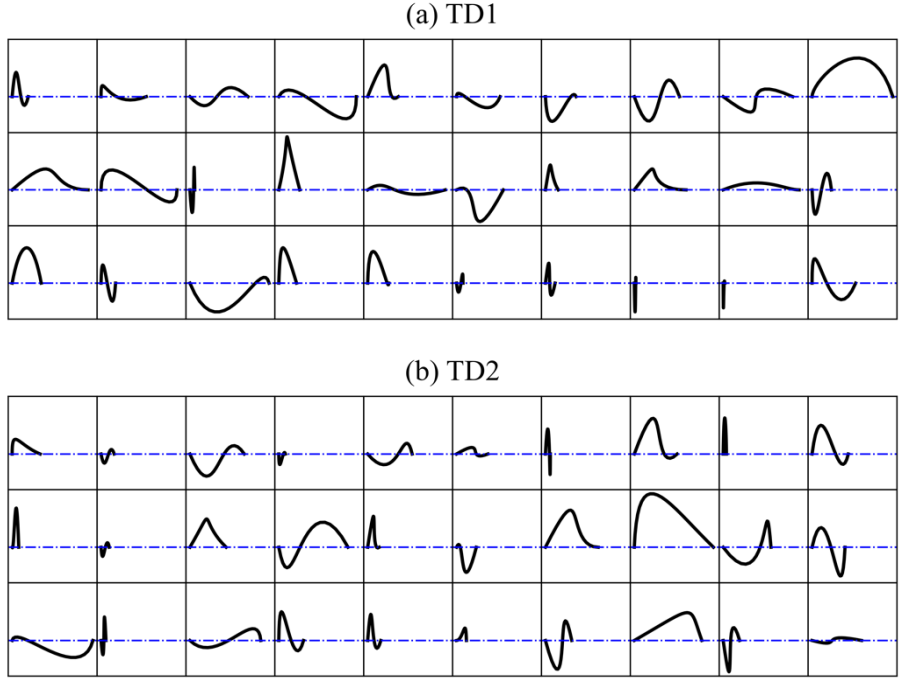

FIG. 8. Randomly generated two independent velocity waveform groups (2*30 samples).

## A. Bayesian optimization results

In the present study, starting from the initial 30 samples, the optimal design parameters of the actuation waveform are sought so as to maximize the multi-objective function $f_\mathrm{I}$ defined by Eq. (12) or $f_\mathrm{II}$ defined by Eq. (13). Since each optimization is started from the two different initial datasets TD1 and TD2, we conduct totally four cases of optimization.

The two multi-objective functions $f_\mathrm{I}$ and $f_\mathrm{II}$ versus a number of iterations during Bayesian optimization are presented in Fig. 9 (a) and (b), respectively. As we explained in the Sec. II C, for the optimization of each multi-objective function, the scale factors $\beta_1 = 0.1$ and $\beta_2 = 0.7$ are determined to make sure that the training datasets leading to a smaller primary droplet and fewer satellites have higher values of the objective functions. As shown in Eqs. (12) and (13), the constant values of 0 and -10 are assigned to the objective function I and II, respectively, when the printer fails to dispense droplets (WD), which are denoted by the blue dash lines in Figs. 9 (a) and (b). Typically, when the inertial force of the ink-thread is not large enough, the droplet cannot overcome the viscous force and the surface tension, and therefore an ink-thread cannot be dispensed from a nozzle. In Fig. 9, successful dispensing including single-droplet dispensing (SD) and multiple-droplet dispensing (MD) are denoted as solid and open markers, respectively. These cases typically have objective functions higher than 0 and -10, respectively. The symbols in the plots are drawn so that their sizes are proportional to the diameter of the resultant droplet, i.e., $d_\mathrm{p}$ and $d_\mathrm{s}$, for the multi-objective functions I and II, respectively. In Fig.

9, the initial 30 samples plotted against blue background are randomly generated training datasets, while the following 120 cases plotted against grey background correspond to waveforms generated through BO algorithm. From the observation of the optimization results, the single and smaller droplet dispensing cases generally have higher objective function values, and this justifies our definition of the multi-objective functions (12) and (13). Both the objective functions are able to find effective waveforms for high-resolution droplet dispensing within only 150 trials, and the minimal droplet diameter is reduced to 32.0% of the nozzle diameter which is quite satisfying comparing to those reported in the previous optimization studies[15,27] for the droplet reduction at similar $Z$ numbers. Comparing between the two objective functions considered in the present study, $f_I$ defined by Eq. (12) is slightly more effective to find better waveforms yielding a small and single droplet (See and compare the numbers of red symbles in Figs. 9 (a) and (b)). We also extend the present optimization with the objective function $f_I$ to the higher $Z$ number of 48. The general trend is similar to the case with the lower $Z$ number, so that the results are listed in Appendix A. For the case of the larger $Z$ number, even a smaller single-droplet with the diameter of 24.9% of the nozzle diameter $d_n$ is achieved (see, in Fig. 16). Comparing to the previous studies summarized in Table IV and Fig. 10, the present BO algorithm together with the proposed multi-objective functions are found to be quite effective for finding waveforms for high-resolution droplet dispensing within the limited trials, especially comparing to the conventional parametric optimization (PO) method.[15,27,28] For PO method, typically tens of the different design points are tried, while for more advanced GA method,[24] more than 1000 samples are produced to optimize droplet dispensing for 4~10 control parameters. In the present study, with a similar dimension of control inputs, BO algorithm can achieve more satisfying optimization results within limited trials of 150, which is very effective for saving experimental time and cost. Moreover, as shown in Fig. 10, the minimum droplet diameter decreases with the $Z$ number, which also indicates the high potential for reducing the droplet size at a high $Z$ number. It should be noted that, however, avoiding satellite generation also beomes more challenging with increasing the $Z$ number.[19]

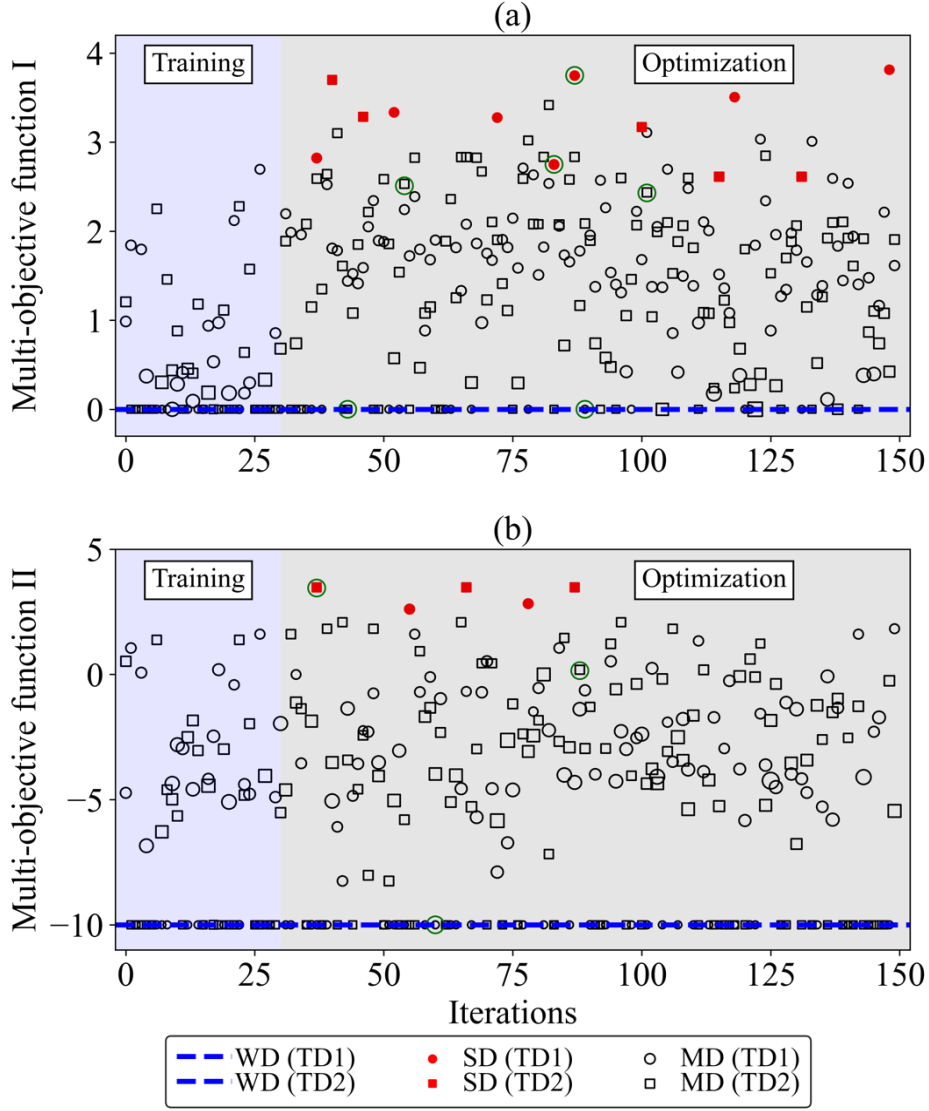

FIG. 9. The values of the multi-objective functions (a) $f_I$ and (b) $f_{II}$ as a function of the number of iterations. Solid markers, open markers and a dash line represent single-droplet (SD), multiple-droplet (MD) and without-droplet (WD) dispensing, respectively. The marker sizes are proportional to generated droplet sizes $d_p$ and $d_s$ for $f_I$ and $f_{II}$, respectively. Initial 30 samples are the training datasets (blue background), and following 120 samples are the trials of Bayesian optimization (grey background). Nine samples marked with green cycles are the selections for the waveform analysis in Sec IV B.

TABLE IV. Summary on the droplet size reduction by waveform tuning.

| Researchers | Waveforms | Optimization methods[+] | Z number | Relative droplet size $(d_s/d_n)*100\%$ |
|---|---|---|---|---|

| Chen and Basaran[15] | Unipolar, bipolar and tripolar | PO | 10 | 45.7% |
| Gan et al.[28] | Unipolar, bipolar, W-type and M-type | PO | 1.4 | 58.5% |
| Aqeel et al.[27] | Unipolar and bipolar | PO | 1.2, 3.6, 10 | 62.0%, 44.9%, 42.7% |
| Snyder et al.[24] | Unipolar, bipolar and tripolar | GA | 51 | 25.6% |
| Present study | Unipolar and bipolar I & II | BO | 9.6, 48 | 32.0%, 24.9% |

[+] PO — Parametric optimization; GA — Genetic algorithm; BO — Bayesian optimization.

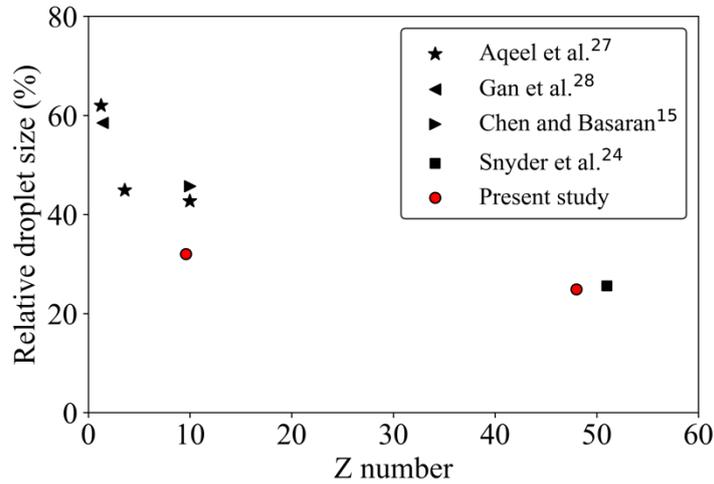

FIG. 10. Droplet size relative to the nozzle diameter achieved in the present and existing studies[15,24,27,28] at different $Z$ numbers.

For DOD printer, the ability to dispense a single-droplet is also referred to as printability which is severely constrained by the ink properties and the nozzle size. In Fig. 11, the printable regions suggested by Derby,[19] Liu and Derby,[22] and Nallan et al.[23] are plotted using the *Re-We-Z* criterion. Here, the nozzle radius is used as the characteristic length in all the cases. The upper limit of the $Z$ number is generally determined by the capability of avoiding satellite droplets. As shown in Fig. 11, the upper limits of $Z$ do not agree well among existing studies, and Snyder et al.[24] reported that single-droplet dispensing can be achieved for a $Z$ number as high as $Z = 51$ if the actuation waveform is optimized. In this study, we consider $Z = 9.6$ and 48, and typical cases with a small primary droplet obtained in the present study are also plotted in Fig. 11. The droplet diameters for $Z = 9.6$ is roughly one third of the nozzle diameter, whereas for $Z = 48$, it is typically one fourth of the nozzle diameter. The solid markers represent single-droplet (SD) dispensing and the open markers are the multiple-droplet (MD) dispensing. Generally, SD samples have relatively lower $Re$ (and $We$) number, which agrees well with the general conclusions in the studies of DOD droplet dispensing.[22,23] However, as shown in Fig. 11, even for $Z = 9.6$, lower $Re$ numbers do not guarantee SD dispensing, and some MD samples can be found even for relatively lower $Re$ number. Therefore, an actuation waveform and associated dispensing processes have a significant impact on printability characteristics.

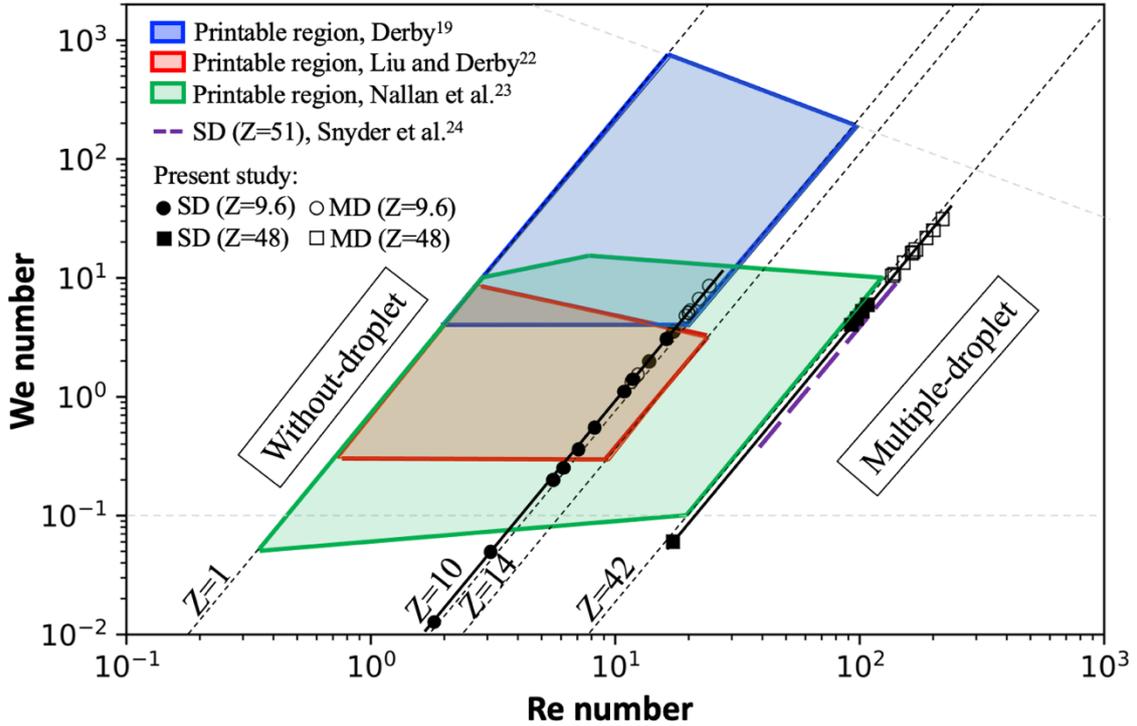

FIG. 11. Printability map for a single droplet using the $Re$-$We$-$Z$ criterion. Cases with single-droplet (SD) and multiple-droplet (MD) dispensing obtained in the present study are shown by solid and open symbols, respectively.

## B. Analysis of obtained waveforms

As shown with red symbols in Figs. 9 (a) and (b), there are several samples which achieve single-droplet dispensing, and therefore result in high values of the objective functions. It turns out that the actuation waveforms of these successful cases can generally be categorized into three different types, i.e., unipolar, and two different types of bipolar profiles. The unipolar waveform is always positive within the actuation period, while the bipolar waveform has both posivie and negative values within the entire actuation period. Hereafter, a waveform that first has a positive value and then a negative value (push-pull) is referred to as bipolar I, whereas its opposite waveform (pull-push) is called biopolar II. Typical successful cases leading to single-droplet dispensing (SD) with unipolar, biplolar I and II waveforms respectively are shown in Figs. 12 (a)-(c). We also add cases with similar waveforms but end up with multi-droplets (MD) and failed (WD) dispensing. The selected samples are identified with green circles in Fig. 9.

For the droplet dispensing using unipolar waveforms shown in Fig. 12 (a), the ink is first ejected from the nozzle and the ink-thread velocity is accelerated, which is corresponding to the rising period of the applied velocity waveforms. For small droplet dispensing, the ejected ink-thread at this period has a length much shorter than nozzle diameter. Then, the ink-thread begins to stretch into a thin thread and the velocity decreases. Viscous dissipation is significant at the initial elongation stage with a sudden contract of ink-thread radius. As shown in Fig. 12 (a), the waveform for WD dispensing has a lower peak velocity comparing to SD and MD dispensing. This suggests that the pushing liquid is not strong enough to overcome the viscous force and the surface tension at the nozzle. Moreover, although the waveforms for the SD and MD cases have the same peak velocity, satellite droplets are eliminated by a rapid decrease of the inlet velocity, which can be explained from the energy budget analysis discussed in Sec. IV D.

For the bipolar I waveforms shown in Fig. 12 (b), the ink-thread evolution during the pushing period is the same as that in unipolar waveforms. Then, a pulling velocity is applied at the inlet boundary. The contraction of ink-thread radius at the nozzle exit pulls air into the nozzle. As shown in Fig. 12 (b), the ink-thread is elongated (see, the ligament inside the nozzle) after the pulling period, which tends to dispense more satellite droplets. In Fig. 12 (b), WD case has a small pushing velocity followed by a large pulling velocity which obviously cannot provide enough inertia to overcome the viscous and surface tension effects. For SD dispensing, a gradual increase in the inlet velocity followed by rapid deceleration is effective for satellite-droplet elimination, which is consistent with the unipolar waveform shown in Fig. 12 (a). We also note that a smaller inlet velocity is required for bipolar I than those for the unipolar and bipolar II respectively shown in Figs. 12 (a) and (c). This is attributed to the elongation of the ink-thread during the pulling period as we discussed above. However, it is also confirmed from Fig. 12 (b) that a longer pulling period could also enhance the generation of satellites.

As for the bipolar II waveforms shown in Fig. 12 (c), a pulling velocity is applied prior to the pushing. When a pulling (negative) velocity is applied at the inlet, the air-liquid interface at the nozzle exit first recedes, and then starts moving outward with a subsequent pushing velocity. For the successful case of SD dispensing, a pulling velocity with a short period is applied before applying pushing. The time period of pushing is similar to those observed in Figs. 12 (a) and (b). However, bipolar II waveform can dispense much smaller droplet size as shown in Fig. 12 (c). It can be explained from the flow focusing effect,[42] where the initial inward deformation of the air-liquid interface makes the ink-thread thinner and a primary droplet smaller. Similar results are also reported in the experiment of Harries et al.[43], where the relative pressure inside the liquid chamber is reduced from the ambient pressure.

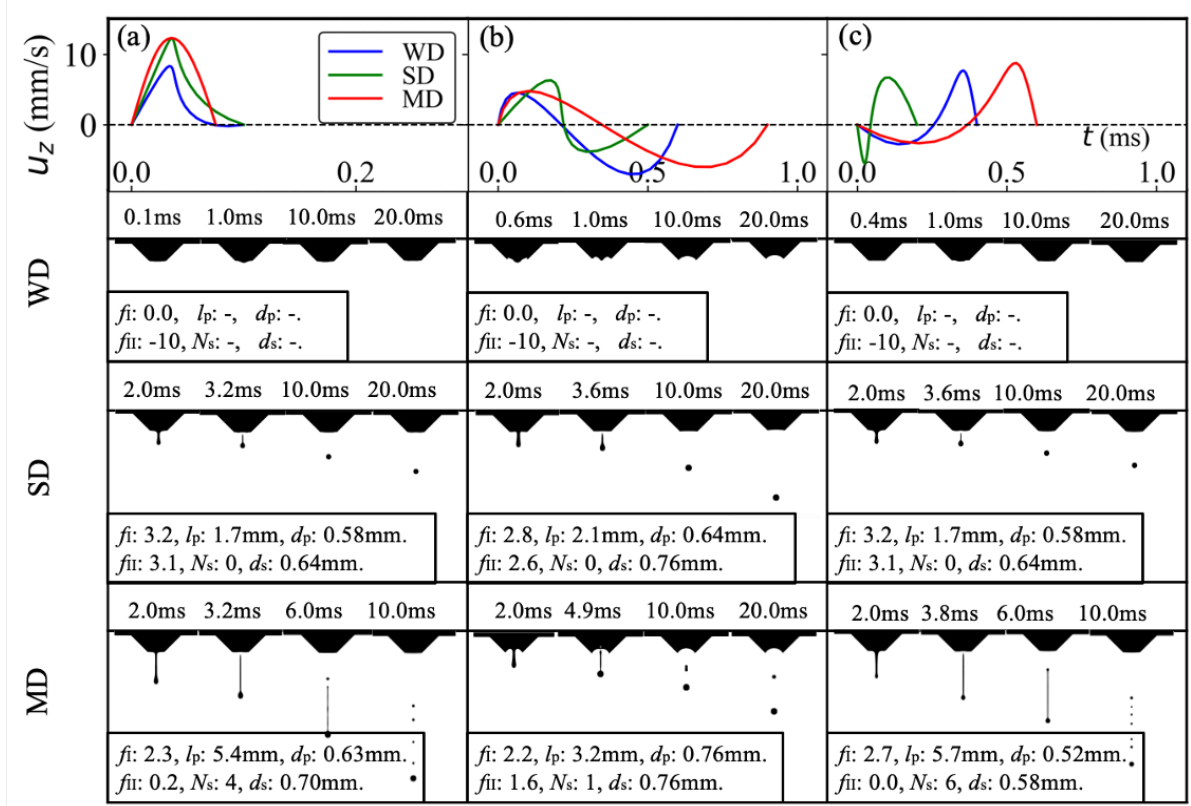

FIG. 12. Comparison of different dispensing with (a) unipolar, (b) bipolar I and (c) bipolar II waveforms. The first row shows the actuation waveforms, and the second to last rows illustrate the droplet morphologies using dispensing images.

### C. Analysis on optimal wave period based on GPR

As shown in Fig. 9, we conducted 150 simulations with different actuation waveforms including the initial random 30 samples for each objective function. Due to the nature of the Gaussian process regression (GPR) model employed in BO, it can provide the estimate of the objective function and its uncertainty for an arbitrary set of the design parameters [$t_1$, $t_2$, $t_3$, $u_{z,1}$, $u_{z,2}$]. Therefore, the present BO algorithm can be used not only for optimizing the design parameters, but also for analyzing how each design variable affects the objective function. Among the five design parameters [$t_1$, $t_2$, $t_3$, $u_{z,1}$, $u_{z,2}$] considered in the present study, $t_3$ determines the period of a waveform, while [$t_1$, $t_2$, $u_{z,1}$, $u_{z,2}$] control the waveform shape and amplitude the characteristics of which have already been discussed in Sec. IV B. Here, we focus on the impacts of the wave period, $t_3$, on the dispensing performance based on the GPR prediction.

Specifically, after training the present GPR model with all the trials, we search the maximum values of the objective functions at each $t_3$ by systematically changing the rest of the four design variables. Note that such a parametric survey based on the GPR model is much cheaper than conducting additional CFD simulations. Then, the maximum values of the estimated objective functions $f_\mathrm{I}$ and $f_\mathrm{II}$ are identified, and their uncertainty is also obtained. The resulting maxima of the estimated objective function for $f_\mathrm{I}$ and $f_\mathrm{II}$ are respectively plotted as a function of the wave period $t_3$ in Figs. 13 (a) and 13 (b), where the uncertainty ranges defined by Eq. (16) are depicted by color bands. In each figure, we plot the results from the two different initial datasets, i.e., TD1 and TD2, with red and blue colors, respectively.

It is confirmed that both the objective functions $f_\mathrm{I}$ and $f_\mathrm{II}$ have peaks around $0.2$ ms $< t_3 < 1.0$ ms, the corresponding region of which is shaded by grey in Figs. 13 (a) and 13 (b). This indicates waves with such a short time period are quite effective in dispensing a single and small droplet. It is also confirmed that the uncertainty around the peak is relatively small, and it is more prominent for $f_\mathrm{I}$. In general, uncertainty becomes smaller when the training data with similar design variables exist. Since BO selects a next sample which is expected to maximize the objective function, it is reasonable that more training data exists around the peak, and therefore the uncertainty is smaller.

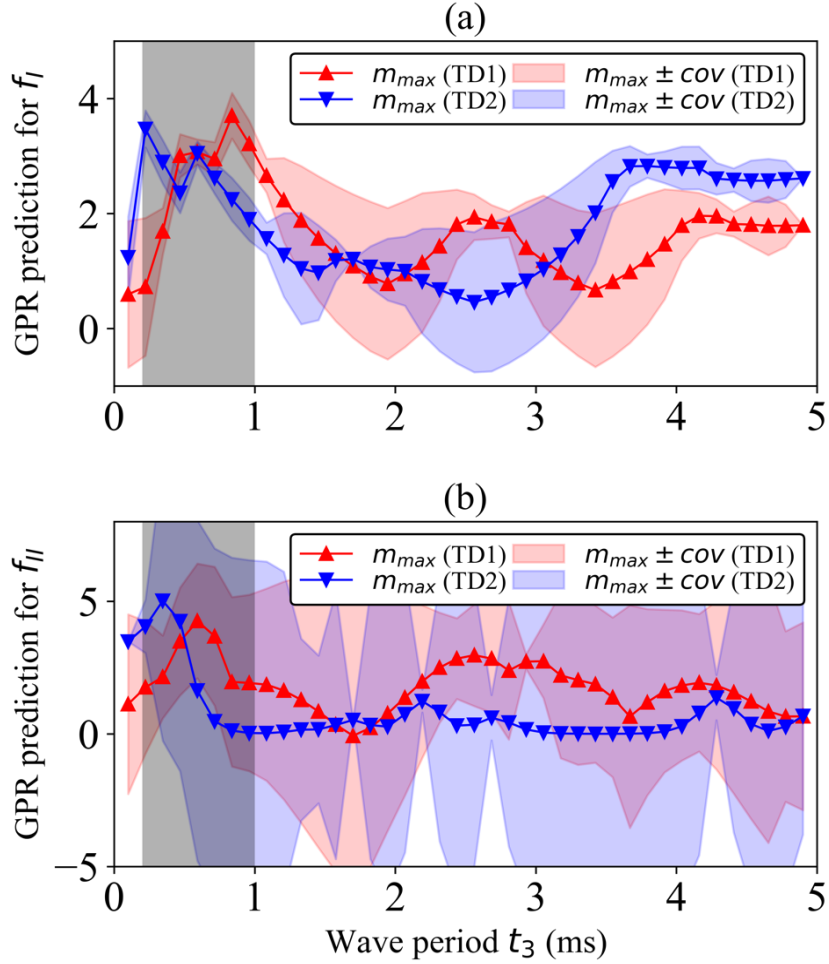

FIG. 13. Maximum predictions with covariance under different wave periods. (a) GPR predictions based on the fitting datasets for $f_\mathrm{I}$, and (b) GPR predictions based on the fitting datasets for $f_\mathrm{II}$. TD1 and TD2 represent two initial training datasets used for the BO searching.

**D. Physical criterion for high-precision droplet dispensing**

As shown in Fig. 12, different types of waveforms can result in single-droplet dispensing. In addition, small differences in the magnitude and the period have significant impacts on the final dispensing results. Here, we aim to develop unified criteria for single-droplet dispensing applicable to all types of waveforms based on the energy budget analysis during the dispensing processes.

At the beginning of the dispensing process, the ink is first ejected from the nozzle with a high speed due to the applied waveform at the inlet. The amount of the ink volume $V_j$ ejected from the nozzle just after applying a waveform can be estimated as:

$$V_j = A_{z=0} \int_0^{t^*} (\bar{u}_z)_{z=0} dt, \tag{22}$$

where $A_{z=0} = \pi d_n^2/4$ and $(\bar{u}_z)_{z=0}$ are the cross-sectional area and the averaged ink velocity in the $z$ direction at the nozzle exit ($z = 0$), respectively, while $t^*$ is the terminal time of the pushing period of the applied waveform at the inlet. Note that the period of applying the inlet velocity $t^*$ is generally much shorter that the entire droplet dispensing process. Therefore, in the following analysis, the flow state at $t = t^*$ is considered as an initial condition.

Assuming that the ink is incompressible, the ink velocity at the nozzle exit during the pushing period can be expressed by using the applied vertical velocity at the inlet ($z = -8.6$) as $(\bar{u}_z)_{z=0} = (A_{z=-8.6}/A_{z=0})(\bar{u}_z)_{z=-8.6}$, where $A_{z=-8.6}$ and $A_{z=0}$ are the cross-sectional areas at the inlet and the nozzle, respectively. Similarly, the kinetic energy of the ejected ink just after the pushing period can be estimated as:

$$E_{k,in} \sim \frac{1}{2} A_{z=0} \rho_l \int_0^{t^*} (\bar{u}_z)_{z=0}^3 dt. \tag{23}$$

It should be noted that the volume and the kinetic energy of the ejected ink just after applying pushing from the inlet defined in Eqs. (22) and (23) can be readily calculated once the applied waveform is given.

When single-droplet dispensing is achieved, the initial kinetic energy of the ejected ink, $E_{k,in}$, defined by Eq. (23) should be balanced by the sum of the kinetic energy $E_{k,d}$ and the surface energy $E_\sigma$ of the droplet after the droplet formation, and also the viscous dissipation $E_\mu$ during the dispensing process. Namely, the following relationship should hold:

$$E_{k,in} = E_{k,d} + E_\sigma + E_\mu. \tag{24}$$

Assuming that the ink volume $V_j$ ejected from the nozzle just after the pushing period is identical to the volume of the dispensed droplet and the velocity of the dispensed droplet is $u_d$, $E_{k,d}$ is given by the following formula:

$$E_{k,d} = \frac{1}{2} \pi \rho V_j u_d^2. \tag{25}$$

Also, assuming that the dispensed droplet is spherical, we obtain:

$$E_\sigma = 4\pi\sigma \left(\frac{3V_j}{4\pi}\right)^{2/3}. \tag{26}$$

As for the viscous dissipation during the dispensing process, assuming that the ink-thread during stretching has a cylindrical shape, we can estimate the viscous dissipation occurs during the entire dispensing process after $t = t^*$ as:

$$E_\mu \sim -3\mu_l V_j \left[\left(1 + \frac{Re_j}{3}\right)\frac{3\mu_l}{2\rho_l}\frac{1}{l_j^2(t^*)} + \frac{u_d}{l_j(t^*)}\right], \tag{27}$$

where $l_j$ is the ink-thread length and and the initial value $l_j(t^*)$ is estimated as $l_j(t^*) = V_j/A_{z=0}$. $Re_j = \rho_l u_{tip}(t^*) l_j(t^*)/\mu_l$ is the Reynold number based on the initial tip velocity $u_{tip}(t^*)$ and length of the ink-thread at $t = t^*$. The initial tip velocity can be calculated from the initial kinetic energy and the ink volume as $u_{tip}^* = \sqrt{2E_{k,in}/\rho_l V_j}$. The detailed derivation of Eq. (27) can be found in Appendix B. It should be noted that only an unknown quantity is the final droplet speed $u_d$, whereas the other variables in Eqs. (25-27) can be evaluated from the initial state of the ejected ink, and thereby calculated from the applied waveform at the inlet.

The energy conservation for a single-droplet dispensing given by Eq. (24) indicates that, in order for a droplet to be dispensed, the initial kinetic energy $E_{k,in}$ has to be larger than the sum of $E_\sigma$ and $E_\mu$. Otherwise, the initial momentum of the ink cannot overcome the surface tension and the viscous force, so that no droplet is dispensed. Hence, the criterion for generating a droplet is expressed as

$$E_{k,in} > E_\sigma + E_\mu^0, \tag{28}$$

where $E_\mu^0$ is estimated assuming that $u_d = 0$ in Eq. (27).

Meanwhile, if the initial kinetic energy is larger than the right-hand-side of Eq. (24), the excess energy should be consumed by additional surface energy $E_\sigma$, thereby results in generation of satellite droplets. Therefore, another criterion for generating a single droplet can be given by

$$E_{k,in} < E_d^{max} + E_\sigma + E_\mu^{max}. \tag{29}$$

where, $E_d^{max}$ and $E_\mu^{max}$ are calculated based on the maximum droplet speed $u_d^{max}$. In the present study, $u_d^{max}$ = 0.5 m/s is assumed based on the printable map shown in Fig. 11 which corresponds to the upper limit of *Re* suggested by Liu and Derby.[22]

the samples considered during the present optimization are plotted in the $V_j$-$E_{k,in}$ plane. Note that the two quantities characterize the initial state of the ink ejected from the nozzle and can directly be computed from the applied waveform at the inlet by Eqs. (22) and (23). Black, red and blue symbols correspond to failed, single-droplet and multiple-droplets dispensing cases, respectively. We also plot the iso-lines of $E_{k,in}/(E_\sigma + E_\mu^0)$ and $E_{k,in}/(E_d^{max} + E_\sigma + E_\mu^{max})$. These iso-lines generally have a similar trend. Specifically, for a large ink volume of $V_j$, the iso-lines gradually go up due to the increase of the surface energy $E_\sigma$. On the other hand, they have steep gradients as $V_j$ approaches to zero because of the increase of the viscous dissipation. It can be seen that the iso-line of $E_{k,in}/(E_\sigma + E_\mu^0) \approx 1$ explains well the boundary of failed and successful dispensing cases, while $E_{k,in}/(E_d^{max} + E_\sigma + E_\mu^{max}) \approx 1$ can be used as a criterion for multiple-droplet dispensing. The present results indicate that the initial ink velocity and ink volume are key factors in determining the final dispensing results.

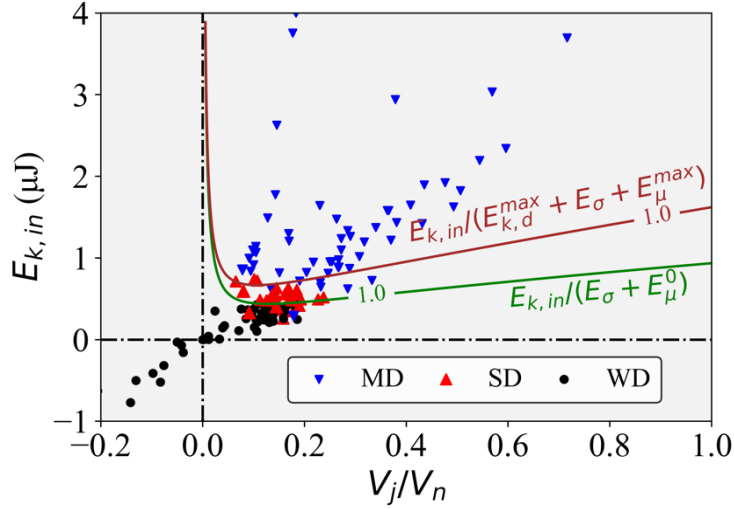

FIG. 14. Map of the initial kinetic energy $E_{k,in}$ and the initial ejected ink volume $V_j$. $V_n$ is the spherical droplet volume with a dimeter of $d_n$. The green and red contours represent the iso-lines of $E_{k,in}/(E_\sigma + E_\mu^0)$ and $E_{k,in}/(E_d^{max} + E_\sigma + E_\mu^{max})$.

## V. CONCLUSIONS

In this study, we develop a simulation code for solving droplet dispensing processes in inkjet printing by applying a multiphase solver, *interFoam*, originally integrated in an open-source CFD code, OpenFOAM. We consider the inlet velocity of the liquid chamber upstream of the injecting nozzle as a control input. By optimizing the time evolution of the inlet velocity, we aim to achieve a single and small droplet dispensing.

First, we validate the current simulation code by comparing the obtained results with existing experimental and numerical data under the same printing condition.[35] Then, we optimize the temporal waveform of the inlet velocity. Specifically, we propose two different multi-objective functions for dispensing a single and small droplet. The first multi-objective function is based on the morphology of the liquid-thread at the pinch-off, while the second one is based on the final dispensing results after the pinch-off. The temporal waveform of the inlet velocity is represented by a cubic *Bezier* curve with four control points, the locations of which are optimized by Bayesian optimization (BO) algorithm.

It is found that both the objective functions successfully find effective waveforms within relatively small trials of 150, which is quite small considering that the number of the design variables is five. A more detailed comparison between the two objective functions shows that the first objective function based on the morphology of the liquid thread at the pinch-off results in more successful cases. The unique feature of the first objective function is that it is defined based on the ratio of the liquid-thread length $l_p$ and the droplet diameter $d_p$ at the pinch-off. This indicates that it is important to take the process of droplet dispensing into consideration when setting the objective function.

We consider two different printing conditions with Z = 9.6 and 48, and it is demonstrated that dispensing of a single and small droplet can successfully be achieved in both cases by using the present optimization strategies. Specifically, a single droplet with a diameter of 32.0% and 24.9% of the nozzle diameter is achieved for Z = 9.6 and 48, respectively. These values are generally smaller than those achieved in existing studies for similar Z numbers.

The waveforms obtained in the present optimization are generally categorized into unipolar, and two different types of bipolar profiles, i.e., bipolar I and II, which apply push-and-pull and pull-and-push actuations, respectively. A common feature of unipolar and bipolar I profiles is that the inlet velocity is increased gradually prior to rapid deceleration. In the case of bipolar II profile, a negative velocity (pulling) before positive velocity (pushing) is effective in reducing the size of a primary droplet.

The prediction from Gaussian process regression (GPR) model indicates that a short wave period is effective for droplet size reduction. Moreover, based on the energy budget analysis, new criteria of an applied waveform for achieving single-droplet dispensing are proposed. The proposed criteria consist of two conditions. Namely, the initial kinetic energy of the ink thread should be larger than the sum of the surface energy of the ejected droplet and the viscous dissipation during the dispensing processes. Meanwhile, the initial kinetic energy should be smaller than the sum of the viscous dissipation and the surface and kinetic energy for a single droplet. Otherwise, the excess initial kinetic energy will create satellite droplets. We demonstrate that the proposed criteria explain all the training samples obtained in the present optimization.

There are several key advantages of the present approach. First, by introducing a *Bezier* curve, complex waveforms can be generated with a relatively small number of design variables. In addition, the sample-efficient Bayesian optimization allows to find the optimal set of design parameters effectively by learning from existing training datasets. Finally, and most importantly, our method takes a black-box approach, so that it can be applied to any system where an actuation waveform determines the final results. For example, the control input can be a voltage signal applied to an actuator, and the present optimization method can easily be extended to experiments. This should be considered in future study.

**ACKNOWLEDGEMENTS**

One of the authors Hanzhi Wang would like to gratefully acknowledge the financial support from JST SPRING, Grant Number JPMJSP2108.

**DATA AVAILABILITY**

The data that support the findings of this study are available from the corresponding author upon reasonable request.

**APPENDIX A: OPTIMIZATION RESULTS FOR $Z = 48$**

Here, we show the optimization results for $Z = 48$. In Fig. 15, the objective function I versus a number of iterations is presented. As for the initial training dataset, TD1 is used, and the first object function, $f_I$, is considered. The optimization conditions are the same as those reported in Fig. 9 (a). Although successful cases with single-droplet dispensing shown with red symbols are fewer than those at the lower $Z$ shown in Fig. 9 (a), it can be confirmed that single-droplet dispensing can be achieved even in the high $Z$ number. Specifically, a single droplet with a radius of 24.9% of the nozzle radius is achieved, which is much smaller than the results optimized by using parametric study[15,27,28] and genetic algorithm[24].

The waveform resulting in the minimum single droplet is shown in Fig. 16. It can be seen that the obtained waveform is similar to the bipolar I profile presented in Fig. 12 (b). This suggests that the optimal waveforms at lower $Z$ numbers can also be effective at high $Z$ numbers.

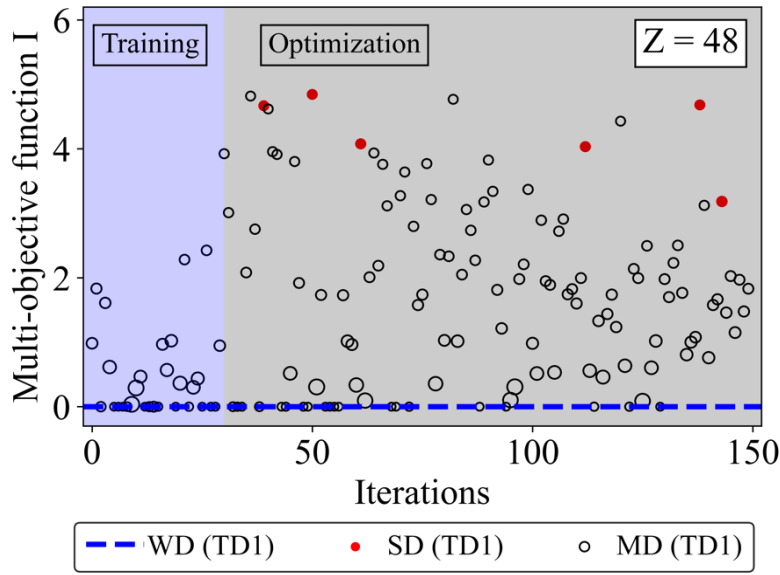

FIG. 15. Objective function vs. iterations for the optimization case of $Z = 48$. Initial 30 samples are the training dataset (TD1) and the following 120 samples are the trials by Bayesian optimization.

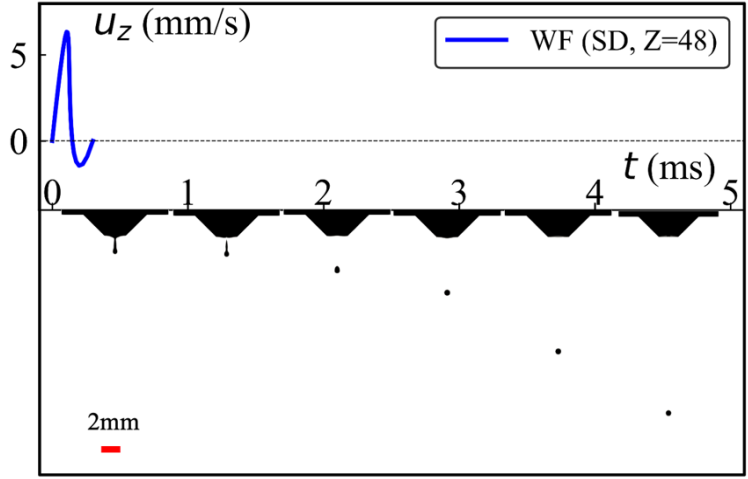

FIG. 16. Actuation waveform (upper) and dispensing images (bottom) for single-droplet (SD) dispensing with $Z = 48$. Images are corresponding to dispensing stages at 2.0, 2.4, 6.0, 10.0, 20.0 and 30.0 ms.

**APPENDIX B: VISCOUS DISSIPATION DURING INK JETTING**

Here, we derive viscous dissipation during the droplet dispensing process. For simplicity, we take the assumptions as follows:

(1) The shape of the ink-thread outside the nozzle is estimated as a cylinder,[44] which has a radius of $r_j(t)$ and a length of $l_j(t)$. Considering that the ink-thread volume $V_j$ is not changing during ink stretching, we get:

$$V_j = \pi r_j^2(t) l_j(t). \tag{30}$$

(2) Assuming that the velocity is uniform in the $r$ direction, the flow in the ink-thread is approximated as a plug flow, $u_z(z,r,t) \sim u_z(z,t)$.[45] In this case, the evolution of the ink-thread radius is governed by the following continuity equation:[46,47]

$$\frac{\partial r_j^2}{\partial t} + \frac{\partial (u_z r_j^2)}{\partial z} = 0. \tag{31}$$

Substituting equation (30) into (31) and keeping in mind that tip velocity $u_{tip}$ is equal to the time derivative of the ink-thread length, i.e., $u_{tip} = \frac{dl_j}{dt}$, we obtain:

$$\frac{\partial u_z}{\partial z} = \frac{u_{tip}}{l_j}. \tag{32}$$

(3) For small droplet dispensing, the ink-thread length at $t = t^*$ typically has a scale of $l_j(t^*) \sim O(10^{-4})$. Even the elongated ink-thread at pinch-off point ($t = t_p$) has a short length scale of $l_j$ ($t_p$) ~ $O(10^{-3})$. Therefore, we approximate the second spatial derivative of velocity with a linearized equation:

$$\frac{\partial^2 u_z}{\partial z^2} \sim -\frac{u_{tip}}{l_j^2}. \tag{33}$$

In order to estimate the viscous dissipation during the dispensing process, we consider the following simplified momentum equation in the $z$ direction:

$$\frac{\partial u_z}{\partial t} + u_z \frac{\partial u_z}{\partial z} = 3\frac{\mu_l}{\rho_l}\frac{\partial}{\partial z}\left(\frac{\partial u_z}{\partial z}\right), \tag{34}$$

where the right-hand-side is the viscosity term for an axisymmetric extensional flow and $3\mu$ is the Trouton viscosity.[48] From Eqs. (32) and (33), we can obtain a linearized momentum equation at the tip location as follows:

$$\frac{\partial u_{tip}}{\partial t} = -\left(\frac{3\mu_l}{\rho_l l_j} + u_{tip}\right)\frac{u_{tip}}{l_j} = -\left(1 + \frac{Re_j}{3}\right)\frac{3\mu_l}{\rho_l l_j}\frac{u_{tip}}{l_j}, \tag{35}$$

where $Re_j$ is the local Reynolds number based on the ink-thread length $l_j$ and the tip velocity $u_{tip}$. During the ink-thread stretching process, $l_j$ increases, whereas $u_{tip}$ decreases due to the viscous force and the surface tension. From our simulations results (not shown here), we confirmed that $Re_j$ does not significantly change during the dispensing process, and therefore we assume that $Re_j$ is constant and equal to its initial value. Moreover, according to Eq. (35), the deceleration of the tip velocity mainly happens at the initial stage of elongation. Therefore, $Re_j$ can be approximated using the initial ink-thread length $l_j(t^*)$ and initial tip velocity $u_{tip}(t^*)$:

$$Re_j = \frac{\rho_l l_j u_{tip}}{\mu_l} \sim \frac{\rho_l l_j(t^*) u_{tip}(t^*)}{\mu_l}. \tag{36}$$

During the ink-thread streching, the viscous dissipation $D_\mu$ is computed from Eqs. (30) and (32) as follows:[45]

$$D_\mu = 3\pi\mu_l \int_0^{l_j} \left(r_j \frac{\partial u_z}{\partial z}\right)^2 dz = 3\pi\mu_l \left(r_j \frac{u_{tip}}{l_j}\right)^2 l_j. \tag{37}$$

By integrating the above function over time, the total viscous energy dissipation before pinch-off ($t = t_p$) is then calculated as:

$$E_\mu = 3\pi\mu_l r_j^2 l_j \int_{t^*}^{t_p} \left(\frac{u_{tip}}{l_j}\right)^2 dt = 3\mu_l V_j \int_{t^*}^{t_p} \left(u_{tip}(t^*) + \int_{t^*}^{t} \frac{\partial u_{tip}}{\partial t} dt\right)\frac{u_{tip}}{l_j^2} dt, \tag{38}$$

where Eq. (35) is used to estimate the deceleration of the tip velocity. By integrating Eq. (35) over time, we arrive at:

$$\int_{t^*}^{t} \frac{\partial u_{tip}}{\partial t} dt = -\left(1 + \frac{Re_j}{3}\right) \frac{3\mu_l}{\rho_l} \left(\frac{1}{l_j(t^*)} - \frac{1}{l_j(t)}\right). \tag{39}$$

Now, considering the pinch-off point which has a tip velocity equal to the droplet velocity $u_d$, we integrate Eq. (39) from $t^*$ to $t_p$ to obtain the following formula:

$$u_{tip}^* - u_d = -\int_{t^*}^{t_p} \frac{\partial u_{tip}}{\partial t} dt = \left(1 + \frac{Re_j}{3}\right) \frac{3\mu_l}{\rho_l} \left(\frac{1}{l_j(t^*)} - \frac{1}{l_j(t_p)}\right). \tag{40}$$

The second term in the second bracket on the right-hand-side can be neglected, since the initial ligament length $l_j(t^*)$ is quite small. Then, substituting Eqs. (39) and (40) into Eq. (38), we get:

$$E_\mu = 3\mu_l V_j \int_{t^*}^{t_p} \left(\left(1 + \frac{Re_j}{3}\right) \frac{3\mu_l}{\rho_l} \frac{1}{l_j(t)} + u_d\right) \frac{u_{tip}}{l_j^2} dt. \tag{41}$$

Integrating the above equation and neglecting function terms relating to ink-thread length at pinch-off point $l_j(t_p)$, we get the function:

$$E_\mu = 3\mu_l V_j \left[\left(1 + \frac{Re_j}{3}\right) \frac{3\mu_l}{2\rho_l l_j^2(t^*)} + \frac{u_d}{l_j(t^*)}\right], \tag{42}$$

where the initial dispensing parameters $l_j(t^*)$ and $Re_j$ can be estimated from the applied waveforms using Eqs. (22) and (23). Here, $u_{tip}(t^*)$ in the definition (36) for $Re_j$ can be calculated from the initial kinetic energy $E_{k,in}$ and initial volume $V_j$ of the ink thread as follows:

$$u_{tip}(t^*) = \left(\frac{2E_{k,in}}{\rho V_j}\right)^{\frac{1}{2}}. \tag{43}$$

In the final form of $E_\mu$ given by Eq. (42), only the velocity $u_d$ of the dispensed droplet is unknown, but the other variables $V_j$, $l_j(t^*)$ and $Re_j$ are determined from the applied waveform at the inlet.